\documentclass[conference,a4paper,9pt]{IEEEtran}
\usepackage
[
        left=1.3cm,
        right=1.3cm,
        top=3cm,
        bottom=2.6cm,
]
{geometry}

\usepackage{amsmath,amsthm,amsfonts,amssymb}
\usepackage{mathrsfs}
\usepackage{mathtools}
\usepackage[english]{babel}
\usepackage{float}
\usepackage[pdftex]{graphicx}
\usepackage{epstopdf}
\usepackage{multirow}

\usepackage{graphicx}
\usepackage{balance}
\usepackage{microtype}
\usepackage[caption=false,font=footnotesize]{subfig}
\usepackage{cite}
\usepackage{comment}
\usepackage{stackengine}
\def\delequal{\mathrel{\ensurestackMath{\stackon[1pt]{=}{\scriptstyle\Delta}}}}

\newcommand{\dddag}{%
  \mathbin{\vbox{\offinterlineskip\ialign{%
    \hfil##\hfil\cr
    \small{$\dagger$}\cr
    \noalign{\kern-0.6ex}
    \small{$\ddagger$}\cr
}}}}

\newcommand{\ddddag}{%
  \mathbin{\vbox{\offinterlineskip\ialign{%
    \hfil##\hfil\cr
    \small{$\ddagger$}\cr
    \noalign{\kern-0.1ex}
    \small{$\ddagger$}\cr
}}}}

\begin{document}

\title{Group-Theoretic Wideband Radar Waveform Design\vspace{-24pt}} 
	
\author{\IEEEauthorblockN{Kumar Vijay Mishra$^{\dag}$, Samuel Pinilla$^{\ddag}$, Ali Pezeshki$^{\dddag}$ 
and A. Robert Calderbank$^{\ddddag}$} 	
        \IEEEauthorblockA{$^{\dag}$United States Army Research Laboratory, Adelphi, MD 20783 USA\\ 
        $^{\ddag}$Tampere University, 33100 Tampere, Finland\vspace{-4pt}\\ 
        $^{\dddag}$Colorado State University, Fort Collins, CO 80523 USA\\
        ${\ddddag}$Duke University, Durham, NC 27708 USA
        }
	}

\maketitle

\begin{abstract}
    We investigate the theory of affine groups in the context of designing radar waveforms that obey the desired \textit{wideband ambiguity function} (WAF). The WAF is obtained by correlating the signal with its time-dilated, Doppler-shifted, and delayed replicas. We consider the WAF definition as a coefficient function of the unitary representation of the group $a\cdot x + b$. This is essentially an algebraic problem applied to the radar waveform design. Prior works on this subject largely analyzed narrow-band ambiguity functions. Here, we show that when the underlying wideband signal of interest is a pulse or pulse train, a tight frame can be built to design that waveform. Specifically, we design the radar signals by minimizing the ratio of bounding constants of the frame in order to obtain lower sidelobes in the WAF. This minimization is performed by building a codebook based on difference sets in order to achieve the \textit{Welch bound}. We show that the tight frame so obtained is connected with the wavelet transform that defines the WAF.
\end{abstract}


\section{Introduction}
The analytical theory of radar signal processing employs cross-correlation technique for target detection and parameter estimation \cite{levanon1988radar,peebles1998radar}. The radar's ability to distinguish closely-spaced targets is completely specified by the \textit{ambiguity function} (AF) of its transmit waveform. Through the AF, the transmit waveform enters into the performance analyses related to detection, target parameter accuracy, and resolution of multiple closely-spaced targets. The AF was first introduced by Ville \cite{ville1948theorie} and its significance as a signal design metric in the mathematical radar theory is credited to Woodward \cite{woodward1965probability,woodward1967radar}, later expounded in detail by Siebert \cite{siebert1956radar}. The AF is not uniquely defined, including in the works of Woodward, which focused on narrowband waveforms \cite{peebles1998radar,george2010implementation,pinilla2021banraw,pinilla2021wavemax}. 

However, many radar applications such as for ground penetration \cite{giovanneschi2019dictionary}, synthetic apertures \cite{vouras2022overview}, and vehicular sensing systems \cite{mishra2019toward} employ wideband waveforms that yield higher range resolution and improved interference suppression. Unlike the narrowband AF (NAF), the Doppler effect on the envelopes of transmit signal reflected off the targets is not constant across the bandwidth. This led to several generalizations of Woodward's definition to wideband ambiguity function (WAF) \cite{speiser1967wide,kelly1965matched,rihaczek1967delay}. In this paper, we focus on such wideband interpretations of AF.

There is a large body of research on using the AF as an aid to select suitable radar waveforms. An AF equal to zero except at one point is ideal for detection tasks \cite[Chapter 3]{jankiraman2007design}. But, in this case, the probability of the target lying within the response region would be near zero \cite{rihaczek1996principles}. In the absence of an ideal AF, significant theoretical efforts have been devoted to the problem of finding functions of delay and Doppler that are not only realizable as AFs but also have suitable radar performance. The resulting procedure of \textit{waveform design} has traditionally focused on achieving AFs that are \textit{thumbtack} with a sharp central spike and low sidelobes in the delay-Doppler plane. The inverse problem - given an AF, design a signal that yields it - is traced back to the 1970 work by Rudolf de Buda. This study claimed that when the NAF is bounded by a \textit{Hermite function}, then the signal is also a Hermite function where the polynomial is found from its AF by comparing coefficients \cite{de1970signals}. Later works such as \cite{jaming2010phase} extended de Buda's results by showing that the bounding assumption over the radar AF is removed to uniquely identify (up to trivial ambiguities) a Hermite function and \textit{rectangular pulse trains}. More recently, phase retrieval techniques have been applied to estimate a band-limited signal from its NAF \cite{pinilla2021banraw}. 

The interest in studying both NAF and WAF in a unified framework resulted in analyzing these functions from the perspective of a group representation theory \cite{chaiyasena1990wavelet}. The NAF and WAF are coefficients of the unitary representations of their respective groups. For example, for the NAF, the delayed and Doppler-shifted replicas of a signal are obtained by a unitary operator that is a member of a Heisenberg group \cite{howard2006finite}. On the other hand, the time dilated and delayed replicas result from the transmit signal through a unitary operator that comes from an affine group. The wideband cross-AFs are, therefore, affine wavelet transforms \cite{chaiyasena1990wavelet}.

In this paper, we investigate group-theoretic formulations to design signals that obey a given WAF. Similar prior works were largely limited to analyzing the NAF. In \cite{howard2006finite}, using the theory developed in \cite{calderbank1997Z4} for the extraspecial 2-group, it was shown that finite Heisenberg-Weyl groups  provide a unifying framework for several sequences to design narrowband phase-coded radar waveforms. Later, this was expanded to design complementary sequences in \cite{pezeshki2008doppler,dang2020coordinating}. Initial investigations into the group representation theory for WAF appeared in \cite{auslander1990wide,chaiyasena1990wavelet,chaiyasena1992wavelet}. This was expanded to include Mellin transform in \cite{shenoy1995wide}. Nearly all of the aforementioned works did not examine the design of wideband waveforms from a given WAF using group-theoretic tools. In this work, we build upon the unitary representation of WAF to design complementary sequences. In particular, we exploit the seminal work of \cite{daubechies1988time,daubechies1990wavelet} to build a tight frame, whose bounding constants are then utilized to obtain low WAF sidelobes. We show that the resulting tight frame is related to the wavelet transform that defines the WAF.

The rest of the paper is organized as follows. In the next section, we introduce the definitions of WAF. We then make a connection with affine groups in Section~\ref{sec:grp}. We analytically design the signal using this theory in Section~\ref{sec:numexp} before concluding in Section~\ref{sec:summ}. Throughout the paper, for reference, the NAF of an uncoded pulse $w(t)$ with duration $T_{c}$ and unit energy
$
\int_{-\frac{T_{c}}{2}}^{\frac{T_{c}}{2}}|w(t)|^{2} d t=1
$ 
is the inner product of the transmit waveform $w(t)$ and its time-delayed/frequency-shifted version: 
\begin{align}
\psi_{w}(\tau, \nu)=\int_{-\frac{T_{c}}{2}}^{\frac{T_{c}}{2}} w(t) w^{*}(t-\tau) e^{-j \nu t} d t,
\end{align}
where $(\tau, \nu)$ are the delay-Doppler coordinates.

\section{WAF of Radar Pulses}
As a traveling wavefield reflects off a moving target, the field either expands or compresses in time because of the movement of the target. When a narrowband waveform is transmitted, this compressive effect is ignored for the waveform's complex envelope and only considered for the carrier. The WAF for which this approximation is not permissible is 
\begin{align}
\chi_{w}^{1}(\tau, \nu)=\sqrt{\gamma}\int_{-\frac{T_{c}}{2}}^{\frac{T_{c}}{2}} w(t) w^{*}(\gamma(t-\tau)) e^{-j \nu t} d t,
\label{eq:WAF}
\end{align}
where the factor $\gamma=1+f_{v} / f_{c}$ accounts for the stretching/compressing in time of the reflected signal (Doppler factor). Assume $f_{0}$ to be the central frequency of the spectra $W(\nu)$. Denote 
\begin{align*}
    \Delta f &= \nu - f_{0}, & \textit{ frequency deviation}, \\
    \delta &= \frac{2v}{c+v}, & \textit{the relative velocity}.
\end{align*}
Then,
\begin{align}
    w(\gamma(t-\tau)) &= e^{-2\pi j f_{0}\delta(t-\tau)}\nonumber\\
    &\times \int_{-\frac{T_{c}}{2}}^{\frac{T_{c}}{2}} W(\nu)e^{2\pi j \nu(t-\tau)}e^{-2\pi j\Delta f \delta(t-\tau)} d\nu.
\end{align}
Observe that there are two factors affecting the signal behaviour in time $e^{-2\pi j f_{0}\delta(t-\tau)}$ and $e^{-2\pi j\Delta f \delta(t-\tau)}$. The first shifts the signal spectra; the second accounts for the stretching/compressing in time of the reflected signal. If the target-source velocity $v\ll c$ then $\delta \ll 1$ and thus $e^{-2\pi j\Delta f \delta(t-\tau)}\approx 1$. In this case, 
\begin{align}
    w(\gamma(t-\tau))=e^{-2\pi j f_{0}(\gamma-1)(t-\tau)}w(t-\tau).
    \label{eq:observation}
\end{align}
It follows from \eqref{eq:observation} that the WAF is, equivalently,
\begin{align}
    \chi_{w}^{1}(\tau, \nu) = \sqrt{\gamma}e^{-2\pi j f_{0}(\gamma-1)\tau}\int_{-\frac{T_{c}}{2}}^{\frac{T_{c}}{2}} w(t) w^{*}(t-\tau) e^{-j \nu t} dt.
    \label{eq:finalWAF}
\end{align}
There is a scalar term in front of the integral to account for a change in amplitude of the reflected signal as it is stretched. The amplitude scaling is necessary for the conservation of energy when the waveform is stretched in time. 

%\subsection{Other forms of WAF}
Note that, like its narrowband counterpart, the definition of WAF is also not unique. The form in \eqref{eq:finalWAF} (hereafter, $\chi_{w}(\tau, \nu)$) is due to Kelly-Wishner \cite{kelly1965matched} and Altes \cite{altes1970methods,altes1973some}. A second form appears as wavelets in the works by Daubechies \cite{daubechies1988time}; Auslander and Gertner \cite{auslander1990wide}; Miller \cite{miller1991topics}:
\begin{align}
\chi_{w}^{2}(\tau, \nu)=\frac{1}{\sqrt{\gamma}} \int_{-\frac{T_{c}}{2}}^{\frac{T_{c}}{2}} w(t) w^\ast\left(\frac{t-\tau}{\gamma}\right)e^{-j \nu t} dt.
\end{align}
Speiser \cite{speiser1967wide} and Chaiyasena et al. \cite{chaiyasena1992wavelet} deal with the following WAF, where the dilation is inside the integral:
\begin{align}
\chi_{w}^{3}(\tau, \nu)=\sqrt{\gamma} \int_{-\frac{T_{c}}{2}}^{\frac{T_{c}}{2}} w(t) w^\ast(\gamma t-\tau) e^{-j \nu t} dt.
\end{align}
In \cite{heil1989continuous,fowler1991signal}, the dilation is included as an exponential (similar to a Doppler shift):
\begin{align}
\chi_{w}^{4}(\tau, \nu)=e^{\frac{\nu}{2}} \int_{-\frac{T_{c}}{2}}^{\frac{T_{c}}{2}} w(t) w^\ast\left(e^{-\nu} t-\tau\right) e^{-j \nu t}dt.
\end{align}
In this paper, we concern ourselves with the definition in \eqref{eq:finalWAF}.

%\subsection{Coded Pulse}
Consider a baseband waveform constructed by phase coding translates of $w(t)$ with a unimodular sequence $x[n]$ of length $L$ as
$$
x(t)=\sum_{\ell=0}^{L-1} x[\ell] w\left(t-\ell T_{c}\right),
$$
with the energy 
$$
\begin{aligned}
E_{x} &=\int_{\mathbb{R}}|x(t)|^{2} d t =\left(\sum_{\ell=0}^{L-1} x[\ell]^{2}\right) \int_{-\frac{T_{c}}{2}}^{\frac{T_{c}}{2}}|w(t)|^{2} d t =L.
\end{aligned}
$$
The WAF $\chi_{x}(\tau, \nu)$ of $x(t)$ becomes %at delay-Doppler coordinates $(\tau, \nu)$ is
\begin{align}
\chi_{x}(\tau, \nu)&= \sqrt{\gamma}\int_{-\frac{T_{c}}{2}}^{\frac{T_{c}}{2}} x(t) x^{\ast}(\gamma(t-\tau)) e^{-j \nu t} dt \nonumber\\
&= \sqrt{\gamma}\int_{-\frac{T_{c}}{2}}^{\frac{T_{c}}{2}} \sum_{\ell,\ell^{\prime}=0}^{L-1} x[\ell] w\left(t-\ell T_{c}\right) x^{*}[\ell^{\prime}] \nonumber\\
&\hspace{8em}\times  w^{*}\left(\gamma t-\gamma\tau - \ell^{\prime} T_{c}\right) e^{-j \nu t} dt \nonumber\\
&=\sum_{\ell,k=0}^{L-1} A_{x}^{\ell}\left(k, \nu T_{c}\right) \chi_{w}\left(\tau- \frac{T_{c}(k+ \ell (\gamma-1))}{\gamma} , \nu\right),
\end{align}
where $\displaystyle A_{x}^{\ell}\left(k, \nu T_{c}\right)= x[\ell] x^{*}[\ell-k] e^{-j \nu \ell T_{c}}$.

%\subsection{Extension to Pulse Trains}
Next, we extend WAF to complementary sequence pair \cite{dang2020coordinating}. Denote a binary sequence of length $N$ by $P=\left\{p_{n}\right\}_{n=0}^{N-1}$ and the complement of $p_{n}$ by $\bar{p}_{n}=1-p_{n}$. The $P$-pulse train $x_{P}(t)$ transmitted at a pulse repetition interval $T$ is 
\begin{align}
\label{eq:biPulseTrain}
    x_{P}(t)=\sum_{n=0}^{N-1} p_{n} x(t-n T)+\bar{p}_{n} \tilde{x}(t-n T).
\end{align}
Multiplying \eqref{eq:biPulseTrain} by a discrete-valued real nonnegative sequence $Q=\left\{q_{n}\right\}_{n=0}^{N-1}$, $\left(q_{n} \geq 0\right)$ of length $N$ gives the $Q$-pulse train $x_{Q}(t)$ is 
\begin{align}
    x_{Q}(t)=\sum_{n=0}^{N-1} q_{n}\left[p_{n} x(t-n T)+\bar{p}_{n} \tilde{x}(t-n T)\right].
    \label{eq:xq}
\end{align}

Transmitting $x_{P}(t)$ and correlating the return with $x_{Q}(t)$ yields
\begin{align}
\chi_{P Q}(\tau, \nu) &=\int_{\mathbb{R}} x_{P}(t) x_{Q}^{\ast}(\gamma(t-\tau)) e^{-j \nu t} d t \\
&\approx\frac{1}{\sqrt{\gamma}}\sum_{n=0}^{N-1} q_{n} e^{-j \nu n T}\left[p_{n} \chi_{x}(\tau, \nu)+\bar{p}_{n} \chi_{\tilde{x}}(\tau, \nu)\right],
\label{eq:initWAF}
\end{align}
where range aliases at offset $\pm n T, n=$ $1,2, \ldots, N-1$ are ignored and the last approximation follows from the fact that $\nu \ll 1/T$ so that the phase rotation within one coherent processing interval (CPI) (\emph{slow time}) can be approximated as a constant. Our goal is to obtain the sequences $p_{n}$ and $q_{n}$ from a given WAF. 

\section{Group-Theoretic Connections of WAF}
\label{sec:grp}
The WAF in \eqref{eq:WAF} is related to a representation of the affine or $a\cdot x + b$ group. %We will also show how to relate the WAF with the wavelets and Mellin transform.
%\subsection{The $a\cdot x + b$ Group}
The $a\cdot x + b$ group is a set $\mathbb{R}\setminus \{0\}\times \mathbb{R}$ equipped with the product
\begin{align}
    (a,b) \circ (a^{\prime},b^{\prime}) = (a\cdot a^{\prime}, b + a\cdot b^{\prime}).
\end{align}
The unity of the group is $(1,0)$ because
\begin{align}
    (1,0) \circ (a,b) &= (a,b) \nonumber\\
    (a,b) \circ (1,0) &= (a,b).
\end{align}
The right and left inverse of $(a,b)$ is $\left(\frac{1}{a},-\frac{b}{a}\right)$ since
\begin{align}
    (a,b) \circ  \left(\frac{1}{a},-\frac{b}{a}\right) &= (1,0) \nonumber\\
    \left(\frac{1}{a},-\frac{b}{a}\right)\circ (a,b)  &= (1,0).
\end{align}
The group is non-commutative
\begin{align}
    (a,b) \circ (a^{\prime},b^{\prime}) &= (a\cdot a^{\prime}, b + a\cdot b^{\prime}) \nonumber\\
    (a^{\prime},b^{\prime}) \circ (a,b) &= (a\cdot a^{\prime}, b^{\prime} + a^{\prime}\cdot b).
\end{align}
From the definition of this group, we now introduce unitary representation.

\subsection{Unitary Representation of WAF}
A representation of the group $\mathbb{R}\setminus \{0\}\times \mathbb{R}$ in the space $L^{2}(\mathbb{R})$ is a pair $\{ \mathcal{U},L^{2}(\mathbb{R}) \}$ where $\mathcal{U}$ is a mapping which assigns to every element $(a,b) \in \mathbb{R}\setminus \{0\}\times \mathbb{R}$ a linear mapping $\mathcal{U}_{(a,b)}: L^{2}(\mathbb{R}) \rightarrow L^{2}(\mathbb{R})$ such that
\begin{align}
    \mathcal{U}_{(1,0)} &= 1_{L^{2}(\mathbb{R})} \nonumber \\
    \mathcal{U}_{((a,b) \circ (a^{\prime},b^{\prime}))} &= \mathcal{U}_{(a,b)} \circ \mathcal{U}_{(a^{\prime},b^{\prime})},
\end{align}
for all $(a,b) \in \mathbb{R}\setminus \{0\}\times \mathbb{R}$. In particular, 
\begin{align}
    \mathcal{U}_{(a,b)} \circ \mathcal{U}_{(\frac{1}{a},-\frac{b}{a})} = \mathcal{U}_{(1,0)}.
\end{align}
If $L^{2}(\mathbb{R})$ is a Hilbert space, a linear representation $\{\mathcal{U},L^{2}(\mathbb{R}) \}$ is said to be unitary if the automorphism $\mathcal{U}_{(a,b)}$ forms a unitary operator of $L^{2}(\mathbb{R})$ for all $(a,b) \in \mathbb{R}\setminus \{0\}\times \mathbb{R}$.

Now consider that $\mathcal{U}$ defines a morphism $(a,b)\rightarrow \mathcal{U}_{(a,b)}$ of the group $\mathbb{R}\setminus \{0\}\times \mathbb{R}$ into a unitary group $\underline{\mathcal{U}}(L^{2}(\mathbb{R}))$ of $L^{2}(\mathbb{R})$ operators such that
\begin{align}
    \mathcal{U}_{(a,b)^{-1}}= \left( \mathcal{U}_{(a,b)} \right)^{*}, \forall (a,b)\in \mathbb{R}\setminus \{0\}\times \mathbb{R}.
\end{align}
Unitary is equivalent to
\begin{align}
    \lVert \mathcal{U}_{(a,b)} f \rVert = \lVert f \rVert, 
\end{align}
for all $(a,b)\in \mathbb{R}\setminus \{0\}\times \mathbb{R}; f\in L^{2}(\mathbb{R}),$ or 
\begin{align}
    \langle \mathcal{U}_{(a,b)} f, \mathcal{U}_{(a,b)} g \rangle = \langle f,g \rangle,
\end{align}
for all $(a,b)\in \mathbb{R}\setminus \{0\}\times \mathbb{R}; f,g\in L^{2}(\mathbb{R})$.

Define the translation representation 
\begin{align}
    (T_{(1,s)} f)(t) = f(t-s), (1,s) \in \mathbb{R}\setminus \{0\}\times \mathbb{R}.
\end{align}
Then it follows that the above representation $T$ is unitary and commutative
\begin{align}
    T_{(1,s)\circ (1,s^{\prime})} = T_{(1,s + s^{\prime})} &= T_{(1,s^{\prime}) \circ (1,s)} \nonumber\\
    &=T_{(1,s)}\circ T_{(1,s^{\prime})}.
\end{align}
The unitary of the translation representation follows from
\begin{align}
    \langle T_{(1,s)}h,T_{(1,s)}g \rangle &= \int_{-\infty}^{\infty} h(t-s)g(t-s) dt \nonumber\\
    &=\langle h, g \rangle.
\end{align}

Define the dilation representation by
\begin{align}
    (D_{\lambda,0}h)(t) = \frac{1}{\sqrt{\lvert \lambda \rvert}} h\left( \frac{t}{\lambda}\right),
\end{align}
for all $(\lambda,0)\in \mathbb{R}\setminus \{0\}\times \mathbb{R}$. It is easy to that the above representation $D$ is unitary and commutative
\begin{align}
    D_{(\lambda,0)} \circ D_{(\lambda^{\prime},0)} &= D_{(\lambda,0) \circ (\lambda^{\prime},0)} \nonumber\\
    &=D_{(\lambda \lambda^{\prime},0)},
\end{align}
and
\begin{align}
    D_{(\lambda^{\prime},0)} \circ D_{(\lambda,0)} &= D_{(\lambda^{\prime},0) \circ (\lambda,0)} \nonumber\\
    &=D_{(\lambda \lambda^{\prime},0)}.
\end{align}
The unitary of the dilation representation follows from
\begin{align}
    \langle D_{(\lambda,0)}h,D_{(\lambda,0)}g \rangle &= \int_{-\infty}^{\infty} \frac{1}{\lambda}h\left(\frac{t}{\lambda}\right) g\left(\frac{t}{\lambda}\right) dt \nonumber\\
    &= \int_{-\infty}^{\infty} h(s) g(s) ds = \langle h, g \rangle,
\end{align}
where a change of variables was needed, i.e. $s = \frac{t}{\lambda}$.

Considering the translation and dilation representations we are ready to define the WAF based on $T$ and $D$. To this end, we consider the product of dilation and translation operators in the set of unitary representations
\begin{align}
    DT_{(\lambda,s)} \delequal D_{(\lambda,0)} \circ T_{(1,s)},
\end{align}
such that for any $h\in L^{2}(\mathbb{R})$, we have
\begin{align}
    (DT_{(\lambda,s)} h)(t) &= D_{(\lambda,0)} \circ (T_{(1,s)} h)(t) \nonumber\\
    &= (D_{(\lambda,0)} h)(t-s) \nonumber\\
    &= \frac{1}{\sqrt{\lvert \lambda \rvert}} h\left( \frac{t-s}{\lambda} \right).
\end{align}
Observe that the product between translation and dilation representations is not commutative 
\begin{align}
    ( T_{(1,s)} \circ D_{(\lambda,0)} h)(t) &= \frac{1}{\sqrt{\lvert \lambda \rvert}} h\left( \frac{t-\lambda s}{\lambda} \right) \nonumber\\
    &\neq ( D_{(\lambda,0)} \circ T_{(1,\lambda s)} h)(t).
\end{align}
We now introduce the notation 
\begin{align}
    h^{(\lambda,s)}(t) &\delequal ( T_{(1,s)} \circ D_{(\lambda,0)} h)(t) \nonumber\\
    &\delequal \frac{1}{\sqrt{\lvert \lambda \rvert}} h\left( \frac{t-s}{\lambda} \right).
\end{align}
Thus, the WAF in \eqref{eq:WAF} can be defined based on the following inner product with a $w(t)$ baseband pulse shape with duration limited to a chip interval $T_{c}$ and unit energy
\begin{align}
    &\langle w,w^{(1/\gamma,\tau)} \rangle \nonumber\\
    &=\sqrt{\gamma} \int\limits_{-\frac{T_{c}}{2}}^{\frac{T_{c}}{2}} w(t) w^{*}\left( \gamma (t-\tau) \right) dt \nonumber\\
    &=\underbrace{\sqrt{\gamma} e^{-2\pi j f_{0}(\gamma-1)\tau}\int\limits_{-\frac{T_{c}}{2}}^{\frac{T_{c}}{2}} w(t) w^{*}(t-\tau) e^{-j \nu t} dt}_{\chi_{w}^{1}(\tau, \nu)},
    \label{eq:design}
\end{align}
which the second comes from \eqref{eq:observation} leading to the same WAF as in~\eqref{eq:finalWAF}.

\subsection{Wavelet Expansion and WAF}
The unitary representation of WAF is connected with the wavelet expansion. Define points of the grid from constants $\gamma_{0},\tau_{0}>0$, $\gamma\not =1,0$. The points of the grid are
\begin{align}
    \frac{1}{\gamma_{m}} = \frac{1}{\gamma_{0}^{m}} \text{ and } \tau_{mn} = n\tau_{0}\frac{1}{\gamma^{m}},
    \label{eq:grid}
\end{align}
for $m,n\in \mathbb{Z}$. Define the function on the grid by the discrete translation and dilation operators 
\begin{align}
    h_{mn}(t) &= ( D_{(1/\gamma_{m},0)} \circ T_{(1,\tau_{mn})} h)(t) \nonumber\\
    &=\frac{1}{\sqrt{\lvert \lambda_{0}^{m} \rvert}} h\left( \frac{t}{\lambda_{0}^{m}} -n\tau_{0} \right).
    \label{eq:wavelet}
\end{align}
Observe that \eqref{eq:wavelet} follows the structure of the wavelet case in \cite{daubechies1990wavelet}. The implication is that we can build a frame in order to design the waveforms $x_{P}(t)$, and $x_{Q}(t)$ in \eqref{eq:biPulseTrain} and \eqref{eq:xq}, respectively. 

In order to build a frame based on wavelet expansion select a constant $K>0 $ and define the operator
\begin{align}
    P(u(t)) = I - \frac{2\sum_{m,n} \langle h_{mn} , u\rangle h_{mn}(t)}{K},
\end{align}
where $I$ is the identity operator. Then, from \eqref{eq:biPulseTrain}, the frame $\tilde{h}_{mn}$ is obtained using $\tilde{h}_{0n}$ by applying the dilation operation as
\begin{align}
    \tilde{h}_{mn}(t) &= D_{(1/\gamma_{m},0)}\tilde{h}_{0n}(t) \nonumber\\
    &=\frac{1}{\gamma_{0}^{m/2}}\tilde{h}_{0n}\left(\frac{1}{\gamma_{0}^{m}}t\right),
\end{align}
where %$\tilde{h}_{0n}$ is given by
\begin{align}
   \tilde{h}_{0n}(t) = \left(\frac{2}{K} \sum_{k} P^{k}\right) h_{0n}(t).
\end{align}
Thus, using this expansion we have that the signal $u(t)$ can be represented as \cite{daubechies1990wavelet}
\begin{align}
    u(t) = \sum_{m,n} \langle h_{mn} , u\rangle \tilde{h}_{mn}(t).
    \label{eq:expansion}
\end{align}

Observe that from \eqref{eq:expansion}, we have that the term $\langle h_{mn} , u\rangle $ can be understood as the WAF. Following \eqref{eq:design}, when $w(t) = u(t)$ and $w^{(1/\gamma,\tau)} = h^{(1/\gamma,\tau)}$. Therefore, the wavelet expansion is applicable for designing the waveforms $x_{P}(t)$, and $x_{Q}(t)$. To this end, we use the following isometry that the frame $\tilde{h}_{mn}(t)$ satisfies:
\begin{align}
    A\lVert u(\cdot) \rVert^{2} \leq \sum_{m,n} \lvert \langle h_{mn} , u\rangle \rvert^{2} \leq B \lVert u(\cdot) \rVert^{2},
\end{align}
for some constants $A>0$ and $B<\infty$. To better represent the signal $u(t)$ using \eqref{eq:expansion}, the ratio $B/A$ needs to be small; ideally, $B/A=1$ \cite{daubechies1990wavelet}. Hence, we pursuit the minimization of $B/A$ by properly selecting the sequences $p_{n}$ and $q_{n}$ in \eqref{eq:xq} since they are the only free parameters of the waveforms $x_{P}$ and $x_{Q}$. The analytical estimation of $A$ and $B$ is usually computationally expensive. Therefore, we target achieving the Welch bound. This is equivalent to minimizing the ratio $B/A$ \cite{xia2005achieving}. We propose to design the sequences $p_{n}$ and $q_{n}$ using a strategy relying on difference sets that have been shown to precisely achieve the Welch bound \cite{xia2005achieving}. 

\section{Numerical Example}
\label{sec:numexp}
\begin{figure}[t!]
	\centering
	\includegraphics[width=1.0\linewidth]{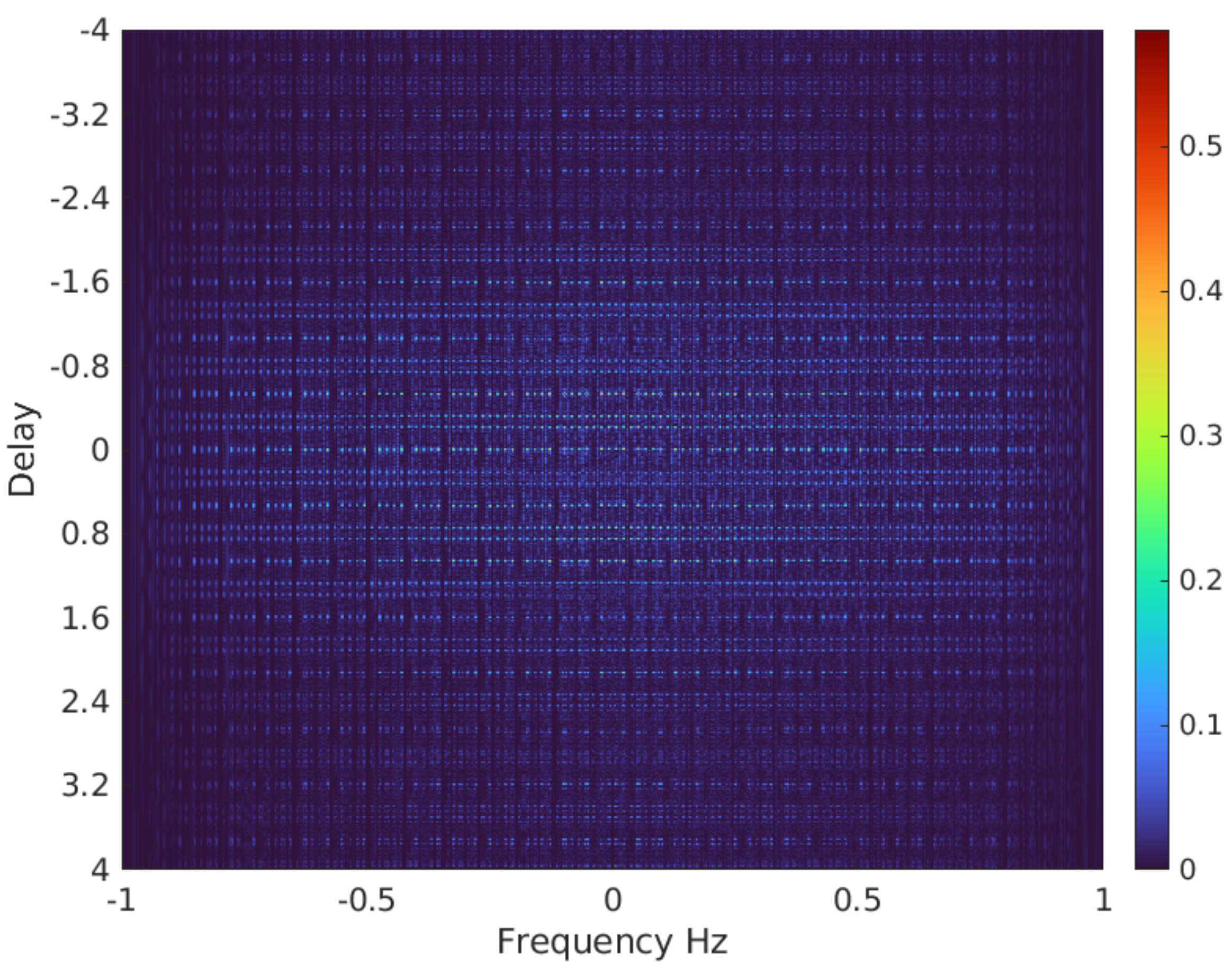}
	\caption{\footnotesize{Wideband cross-AF for the designed signals $x_{P}$ and $x_{Q}$.}}
	\label{fig:orthoini}
\end{figure}
Difference sets have been well studied in the combinatorial design theory \cite{dinitz1992contemporary,aigner2012combinatorial}. They are known to exist for certain cases and a comprehensive repository of difference sets is provided in \cite{list2021}. As an illustration of our approach, we employ the Singer difference sets \cite{singer1938theorem} to design the sequences $p_{n}$ and $q_{n}$ in \eqref{eq:xq}.   

Consider the prime number $q=11$ and a power constant $d=2$. Then, the size of the sequences is directly determined from these values as $$N=\frac{q^{d+1}-1}{q-1}=133.$$ Define $C_{1} = \frac{q^{d}-1}{q-1}$ and $C_{2}=\frac{q^{d-1}-1}{q-1}$. This yields a $(N,C_{1},C_{2})$-difference set. Under this setup, the resultant difference set is \begin{align}
    \mathcal{F} = \{0,  1,  8, 14, 30, 45, 47, 56, 66,106,109,129 \}.
    \label{eq:set}
\end{align}
From \eqref{eq:design}, the WAF in \eqref{eq:initWAF} corresponds to the inner product between $x_{P}$ and $x_{Q}$. The function uses the product $p_{n}q_{n}$ as
\begin{align}
    &\left \lvert \langle x_{P},x_{Q}^{(1/\gamma,\tau)} \rangle \right \rvert \equiv \chi_{P Q}(\tau, \nu) \nonumber\\
    &\approx \frac{1}{\sqrt{\gamma}}\sum_{n=0}^{N-1} q_{n} e^{-j \nu n T}\left[p_{n} \chi_{x}(\tau, \nu)+\bar{p}_{n} \chi_{\tilde{x}}(\tau, \nu)\right] \nonumber\\
    %\left \lvert \langle x_{P},x_{Q}^{(1/\gamma,\tau)} \rangle \right \rvert
    &\leq \frac{1}{\sqrt{\gamma}}\sum_{n=0}^{N-1} q_{n}p_{n}\lvert \chi_{x}(\tau, \nu) \rvert + q_{n}(1-p_{n})\lvert \chi_{\tilde{x}}(\tau, \nu) \rvert.
\end{align}

Therefore, in order to minimize the ratio $B/A$ of the resultant frame applied to $x_{Q}$ (following the strategy explained in the previous section), we need to design $p_{n}q_{n}$. To this end, we set $p_{n}q_{n}=1$ for all $n\in \mathcal{F}$, and $p_{n}q_{n}=0$ for all $n \in \{0,\dots,N \}\setminus \mathcal{F}$. From these constraints, we can determine $(p_{n},q_{n})$ as
\begin{equation}
    (p_{n},q_{n}) = \left\lbrace \begin{array}{ll}
         (1,1) & \text{ if } n \in \mathcal{F} \\
         (n\mod 2,n+1 \mod 2) & \text{ if } n\not \in \mathcal{F}
    \end{array} \right..
\end{equation}
The resultant wideband cross-AF is plotted in Fig.~\ref{fig:orthoini}.

In the context of tight frames in \cite{tropp2008conditioning,mosher2012non}, it has been shown that, if the Welch bound is reached, with overwhelming probability we are able to guarantee the reconstruction. It turns out that difference sets can provide patterns that reach the Welch bound \cite{xia2005achieving}. For the particular scenario we considered, this bound is $\mu = \sqrt{\frac{N-C_{1}}{C_{1}(N-1)}}=\sqrt{\frac{121}{1584}}\approx 0.27$. Therefore, $\frac{B}{A}=1+\mathcal{O}(\mu)$. We remark that this strategy is limited by the calculation of the difference set in \eqref{eq:set}, which is computationally demanding. In the future, we intend to develop a numerical algorithm to estimate new difference sets to produce higher values of $N$.

\section{Summary}
\label{sec:summ}
We analytically demonstrated design procedure a wideband radar waveform using group-theoretic strategy via difference sets. We first mathematically expressed the waveform's WAF as a coefficient function of the unitary representation of the group $a\cdot x + b$. From this representation, we connected the wavelet expansions of frame to the WAF. This allowed us to derive a strategy to design two binary sequences in order to minimize the ratio of the bounding constant of the wavelet frame so constructed. The analytical estimation of these bounds is usually computationally expensive. As an alternative, we targeted achieving the Welch bound using difference sets. Our proposed group-theoretic waveform design guarantees a near-optimal reconstruction. 

\begin{comment}
The $\chi_{P Q}(\tau, \nu)$ is well approximated by
$$
\begin{aligned}
\chi_{P Q}(\tau, \nu) &=\sum_{n=0}^{N-1} q_{n} e^{-j \nu n T} \\
\times & \sum_{k=-(L-1)}^{L-1}\left[p_{n} C_{x}[k]+\bar{p}_{n} C_{y}[k]\right] C_{\Omega}\left(\tau-k T_{c}\right)
\end{aligned}
$$
A key fact to take from $(18)$ is that it is a linear combination of translates of $C_{\Omega}$, and of course that this function has support twice the chip-length about the origin. Accordingly, it is convenient to leave this aside and consider
$$
\begin{array}{l}
\chi_{P Q}(k, \theta) \\
\quad=\sum_{k=-(L-1)}^{L-1}\left[p_{n} C_{x}[k]+\bar{p}_{n} C_{y}[k]\right] \sum_{n=0}^{N-1} q_{n} e^{-j \nu T}
\end{array}
$$
which, after some simple algebraic manipulations, we can write as
$$
\begin{aligned}
\chi_{P Q}(k, \theta)=& \frac{1}{2}\left[C_{x}[k]+C_{y}[k]\right] \sum_{n=0}^{N-1} q_{n} e^{j n \theta} \\
&-\frac{1}{2}\left[C_{x}[k]-C_{y}[k]\right] \sum_{n=0}^{N-1}(-1)^{p_{n}} q_{n} e^{j n \theta}
\end{aligned}
$$
where $\theta=\nu T$ is the relative Doppler shift over a PRI $T$.
\end{comment}

%\clearpage
\bibliographystyle{IEEEtran}
\bibliography{main}

\end{document}